\documentclass[conference]{IEEEtran}
\IEEEoverridecommandlockouts
\usepackage{cite}
\usepackage{amsmath,amssymb,amsfonts}
\usepackage{algorithmic}
\usepackage{graphicx}
\usepackage{textcomp}
\usepackage{xcolor}
\usepackage{float}
\usepackage{varioref}
\usepackage{mathtools}
\DeclarePairedDelimiter{\ceil}{\lceil}{\rceil}
\def\BibTeX{{\rm B\kern-.05em{\sc i\kern-.025em b}\kern-.08em
    T\kern-.1667em\lower.7ex\hbox{E}\kern-.125emX}}
\begin{document}

\title{Moving horizon-based optimal scheduling of EV charging: A power system-cognizant approach}

\author{\IEEEauthorblockN{Nitasha Sahani, Manish Kumar Singh, Chen-Ching Liu}
\IEEEauthorblockA{Emails:\{nitashas,manishks,ccliu\}@vt.edu\\
Bradley Department of Electrical and Computer Engineering, Virginia Tech, Blacksburg, VA 24061, USA}
}

\maketitle

\begin{abstract}
 The rapid escalation in plug-in electric vehicles (PEVs) and their uncoordinated charging patterns pose several challenges in distribution system operation. Some of the undesirable effects include overloading of transformers, rapid voltage fluctuations, and over/under voltages. While this compromises the consumer power quality, it also puts on extra stress on the local voltage control devices. These challenges demand for a well-coordinated and power network-aware charging approach for PEVs in a community. This  paper formulates a real-time electric vehicle charging  scheduling  problem as an mixed-integer linear program (MILP). The problem is to be solved by an aggregator, that provides charging service in a residential community. The proposed formulation maximizes the profit of the aggregator, enhancing the utilization of available infrastructure. With a prior knowledge of load demand and hourly electricity prices, the algorithm uses a moving time horizon optimization approach, allowing the number of vehicles arriving unknown. In this realistic setting, the proposed framework ensures that power system constraints are satisfied and guarantees desired PEV charging level within stipulated time. Numerical tests on a IEEE 13-node feeder system demonstrate the computational and performance superiority of the proposed MILP technique.
 \end{abstract}

\begin{IEEEkeywords}
Electric vehicle, Charging Scheduling, Mixed Integer Linear Programming (MILP), Moving time horizon optimization
\end{IEEEkeywords}

\section{Introduction}
The main advantage of transportation electrification over internal combustion engine vehicles is reduction in fuel consumption. This leads to diminished greenhouse gas emissions, increase in clean energy usage and transportation energy efficiency enhancement. Requirement of plug-in electric vehicle (PEV) charging infrastructure and the impact of PEV load on the present distribution system limits PEV adoption. PEV load demand increase calls for infrastructure addition on generation, transmission, and distribution systems. The upgradation of these infrastructure is often capital intensive and has a long time-lag. However, a meticulously designed charging approach could minimize the infrastructure-upgradation requirements.

Besides increased load demand, other issues related to PEVs like compromised power quality, circuit reliability and the longevity of transformers can be mitigated by coordinated charging of PEVs. The three major concerns related to PEV charging are: \emph{i)} satisfying power system operation constraints; \emph{ii)} meeting customers' PEV charging requirements with guarantees; and \emph{iii)} catering to uncertainty in PEV arrivals over the operation period. 

The centralized controlled charging methods implemented in [1]-[5] focus on utility requirements where grid constraints and minimizing the operational cost were primary goals. Such approaches do not incorporate customer preference and hence do not guarantee complete charging of PEVs by the end of available time. Similarly, a decentralized counterpart for PEV charging technique is formulated meeting the power system limits, but ignoring user preferences [6]. Several works have proposed charging scheduling approaches that meet customer requirements but ignore the modeling of power distribution system. For instance, references [7] and [8] used the centralized charging scheduling approach in which customer's choice and charging profiles were taken into account. Reference [9] proposed centralized strategy with complex communication network to provide PEV charging schedule. A three-level hierarchical framework for decentralized control has been proposed in [10] to improve PEV charging and reduce the grid operation cost. Some studies like [11]-[13] considered both circuit constraints and consumer's choice along with dynamic PEV arrival rate to reduce the charging cost for a fixed charging rate. In [14], the scheme for PEV charging management is based on heuristic algorithm and uses genetic algorithm optimization which also takes into account network constraints, charging requirements and user driving behaviour. The above methods did not tend ensure the charging satisfaction along with grid and customer requirements.

This work proposes a novel computationally tractable algorithm for real-time PEV scheduling based on a moving time horizon setting. Charging schedules are generated and revised periodically based on the actual number of PEVs arriving in real-time. The major contributions of the proposed formulation include:
\begin{enumerate}
\item The number of PEVs arriving in real-time is kept unknown, making the setup realistic.
\item Power system operational constraints are met for all operational periods.
\item Customers are offered different price choices that determine the time-of-return with required charging levels. Despite revisions to the charging schedules in every period, the time-of-return is satisfied for all customers under contract.
\end{enumerate}
The paper is organized as: Section II describes the system model considered; Section III provides the charging scheduling problem formulation and the optimization algorithm; the numerical tests for the proposed method is shown in Section IV. Section V talks about the conclusion and future work. 

\section{System Model}

\subsection{Charging Model Setup}

The PEV charging setup considered in this work comprises of an aggregator providing charging service to a residential community. During the entire operating period, the aggregator is supposed to control the total real-time load of the station such that the power distribution system constraints for the entire feeder are satisfied. This arrangement minimizes the burden on utility for charging schedule and also reduces the dependency on high-end communication technology. It is assumed that the aggregator receives a day-ahead electricity prices, forecasted load, and forecasted generations at all buses. A deterministic model is considered for simplicity. However, it is assumed that the aggregator is not aware of the number of PEV's arriving at different times across the day. Hence, a moving horizon based scheduling algorithm is developed that periodically updates the schedule based on real-time arrival of PEVs at the charging station.

The main novelty of the proposed algorithm is that if a PEV arrives at the charging station and a contract is established, the scheduling process guarantees that the charging commitments are met despite of uncertain future schedule revisions. The charging commitments comprise of two components: \emph{i)} time of return after charging; and \emph{ii)} final charging level of PEV on return. While the final charging level is determined by customers' needs, the time of return is determined as described next. When a PEV arrives with a charging request, the aggregator gives the PEV owner the flexibility of choosing a charging cost option. For every cost option, the charging time required by the aggregator for charging the PEV to its required level will be provided. Based on this information, the PEV owner decides the charging cost. The charging time limit is decided by dividing the required charging power with an average charging rate for each cost option. In this setup, two charging cost options are given to the PEV owner.



Two charging prices $C_1$ (\$/kWh) and $C_2$ (\$/kWh) are considered here. This can be scaled for more number of charging options. Here, $C_1$ is greater than $C_2$. So, the costumers choosing $C_1$ as charging cost have higher priority than that of $C_2$. The reason for a customer opting for $C_1$ is less charging time guaranteed by the aggregator. The PEVs selecting the cost option $C_1$ are grouped as one category and PEVs selecting $C_2$ are grouped as another. The aggregator solves the optimization problem of maximizing its charging profit by generating optimal charging schedule at the beginning of each 1 hour time interval.
  
With new PEVs requesting charging, the aggregator can revise the previous charging schedule facilitating maximal PEV charging at all times, the aggregator can alter the previous charging schedule, based on the number of new PEVs requesting charging. Depending on the available charging capacity, power system load levels, and existing charging commitments from previous hours, the aggregator selects the PEVs from the two groups for providing the charging power. 


\begin{figure}[t]
\centering
\includegraphics[width=1\linewidth]{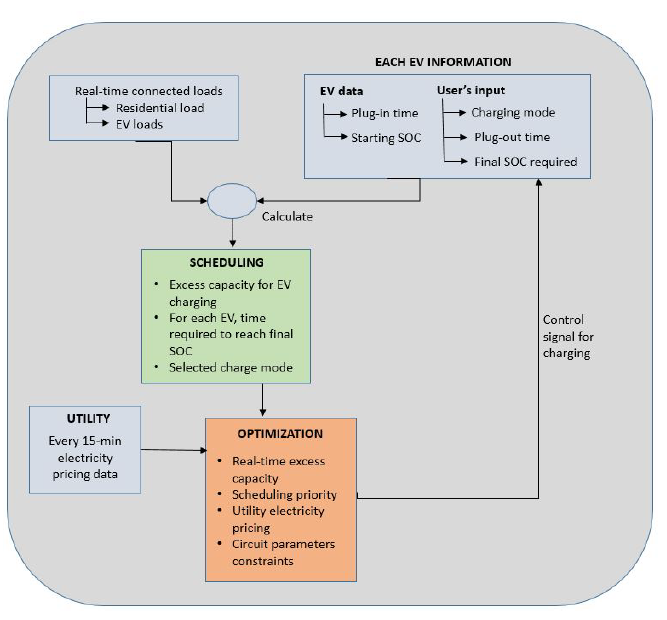}
\caption{Aggregator's Algorithm Setup }\label{Fig:}
\end{figure}

In this setup, PEVs arriving for charging between two intervals are assumed available for charging from the next time step. Also, the charging for PEV is not necessarily continuous, i.e., a PEV might sometimes remain idle. The schematic diagram of the scheduling arrangement is summarized in Figure 1.

\subsection{Residential Load Profile}

In a residential area, peak load hours are usually 6 pm-8 pm. During a weekday evening, residents come home from work and basic residential loads are in use, along with which PEV load is plugged in under uncoordinated charging. This practice leads to higher loading conditions. The off-peak hours in a residential setup is from 11 pm- 6 am and 8 am- 2 pm. So, the primary idea behind scheduling the PEV charging is to spread out the charging start times causing less steep peaking.

\subsection{Electric Vehicles Charging Specifications}
We consider Level 2 charging that uses a 240V AC setup and can be configured for variable charging power of 3.3- 19.2 kW. Depending on the make, model, distance range and affordability, PEVs can have varying range of battery capacities. The charging efficiency ($\eta$) is used as with increase in usage and losses incurred by the charging setup, the battery charged is less than the input grid energy.

\section{Problem Formulation}
In this section, an MILP is formulated that the aggregator solves at the beginning of every time interval and generates a charging schedule for the remaining time intervals. A single-phase radial power distribution system is considered for clarity of the formulation. The developed approach however can be conveniently extended to multi-phase unbalanced systems. 

\subsection{Power System Constraints}\label{AA}
Consider a single phase radial distribution system with $N$ nodes. Let $p_{i,t}$ and $q_{i,t}$ represent the net active and reactive power injection at node $i$ during time $t$. It is assumed that the active and reactive power generations $(p_{i,t}^g, q_{i,t}^g)$, and loads $(p_{i,t}^\ell, q_{i,t}^\ell)$ are known a priori. The nodes hosting the aggregator charging facility will have additional active power load $p_{i,t}^{EV}$. Thus power balance for any network node $i$ entails
\begin{subequations}\label{eq:pq}
	\begin{align}
	p_{i,t}&=p_{i,t}^g-p_{i,t}^\ell-p_{i,t}^{EV},~\forall t\\
	q_{i,t}&=q_{i,t}^g-q_{i,t}^\ell,~\forall t.
	\end{align}
\end{subequations}
Oftentimes there are maximum apparent power limits $\Bar{s}_i$ based on feeder rating, transformer rating or contracted capacity. These constraints are quadratic in general. However, with known reactive power, the quadratic constraints $p_{i,t}^2+q_{i,t}^2 \leq \overline{s}_i^2 $ can be written as linear constraints 
\begin{equation}
    |p_{i,t}| \leq  \sqrt{\overline{s}_i^2- q_{i,t}^2}.
\end{equation}

Given the nodal power injections, the voltages may be determined using the power flow equations. In this formulation the linearized distribution flow (LDF) model is employed for computational tractability [15]. Let $v_{i,t}$ be the squared voltage at node $i$ at time $t$, and $\mathbf{v}_t$ be the $N-1$-length vector collecting all nodal squared voltages other than the substation node voltage $v_0$ for time $t$. Similarly, let $\mathbf{p}_t$ and $\mathbf{q}_t$ be the collection of active and reactive power injections at non-substation nodes, respectively. Then, the linear power flow model dictates
\begin{equation}
\mathbf{v}_T=v_0\mathbf{1} +\mathbf{R}\mathbf{p}_t+\mathbf{X}\mathbf{q}_t
\end{equation}
where $\mathbf{R}$ and $\mathbf{X}$ are derived from the network topology and impedances as detailed in [15]. Next, any stipulated limits on power injections and voltages may be imposed as
\begin{subequations}\label{eq:limits}
	\begin{align}
	\underline{p}\mathbf{1}\leq\mathbf{p}_t\leq\overline{p}\mathbf{1},~\forall~t\\
	\underline{q}\mathbf{1}\leq\mathbf{q}_t\leq\overline{q}\mathbf{1},~\forall~t\\
	\underline{v}\mathbf{1}\leq\mathbf{v}_t\leq\overline{v}\mathbf{1},~\forall~t.
	\end{align}
\end{subequations}
where ($\underline{p},~\underline{q}$ and ($\overline{p},~\overline{q}$) are the limits on active and reactive power respectively. The voltage limits $\underline{v}$ and $\overline{v}$ are the minimum and maximum permissible squared voltages of the distribution system. As a voltage deviation between a distribution transformer and the service voltage is expected, the voltages at the distribution transformers level is maintained within $\pm 3\%$ pu, resulting in the squared voltage limits as [$0.97^2~1.03^2$].

\subsection{PEV Scheduling Constraints}
For a given day, the entire time horizon may be divided into intervals of length $\Delta t$, say $\Delta t=1$~hr. Let us index the intervals as $t=1,2,\cdots,T$. Let $M_k$ denote the number of PEVs that have entered a charging contract by the $k$-th interval and have not yet reached the desired charging level. Denote the number of new PEVs requesting charging at the beginning of $k$-th interval as $n_k$. Thus, at the beginning of $k$-th interval, the aggregator would try to prepare a charging schedule for $N_k:=M_{k-1}+n_k$ over the remaining $T_k=(T-k+1)$ intervals $t=k,k+1,\cdots,T$. An optimal schedule shall first ensure that the charging requirements for the previously contracted $M_{k-1}$ PEVs is fulfilled in the remaining $T_k$ intervals. Next, a subset of the new $n_k$ PEVs shall be selected to enter a charging contract such that the optimal schedule can successfully fulfill their charging requirements. Indexing the $N_k$ PEVs by ${n=1,\dots,N_k}$, let the binary variable $u_n$ denote whether a contract is established with PEV $n$ ($u_n=1$); or otherwise $(u_n=0)$. Since the first $M_{k-1}$ PEVs are already contracted from previous binary instances, we have
\begin{subequations}\label{eq:un}
	\begin{align}
	u_n&=1,\quad \forall n=1,\dots, M_{k-1}\\
	u_n&\in\{0,1\},\quad \forall n=M_{k-1},\dots,N_k.
	\end{align}
\end{subequations}
Note that while $M_k$ and $n_k$ are problem parameters, the decision for establishing a contract $u_n$ is an optimization variable.

Define matrices $\mathbf{D}^{(k)}\in\{0,1\}^{N_k\times T_k}$, and $\mathbf{P}^{(k)}\in\mathbb{R}^{N_k\times T_k}$ to represent the overall schedule prepared at the beginning of $k$-th interval, as detailed next. The binary entry $D_{nt}=1$ represents that PEV $n$ is scheduled to receive a charging power $P_{nt}$ during the $t$-th time interval. Otherwise, entry $D_{nt}=0$ implies PEV $n$ remains idle during interval $t$, and receives no charging. Since only the PEVs entering a contract shall participate in the schedule, the following constraints hold

\begin{subequations}\label{eq:unPD}
	\begin{align}
	D_{nt}&\leq u_n,\quad \forall n,~t\\
	\underline{p}^{EV}\mathbf{D}^{(k)}&\leq \mathbf{P}^{(k)}\leq\overline{p}^{EV}\mathbf{D}^{(k)} 
	\end{align}
\end{subequations}
where $[\underline{p}^{EV},~\overline{p}^{EV}]$ represent the limits on charging power of PEVs. The assumption of common limits $[\underline{p}^{EV},~\overline{p}^{EV}]$ is without loss of generality.

Let $a_n^k$ denote the number of time intervals starting from the $k$-th interval within which PEV $n$ must receive $s_n^k$ units of energy. The values for parameters $a_n^k$ and $s_n^k$ are derived based on the contract established between the aggregator and PEV owner, and is updated after every time interval as detailed in Section III-D. The following constraints impose the requirements for charging times and total charge needed
\begin{subequations}\label{eq:chargetime}
	\begin{align}
	D^{(k)}_{nt}&=0,\quad \forall t>a_n^{(k)},~ \forall n\label{seq:time}\\
	\mathbf{P}^{(k)}\mathbf{1}&=\mathbf{u}\odot \mathbf{s}^{(k)}\label{seq:charge}
	\end{align}
\end{subequations}
where $\mathbf{u}$ and $\mathbf{s}^{(k)}$ collect the contract statuses $u_n$'s and charging requirement $s_n^k$'s for the $N_k$ PEVs, and $\odot$ represents the entry-wise product of the two vectors. Constraint \eqref{seq:time} ensures that in the prepared schedule at the beginning of interval $k$, PEV $n$ receives no charging power after time $a_n^k$. Constraint \eqref{seq:charge} ensures that the row-sum of $\mathbf{P}^k$, representing the total charge received by PEV $n$, considering $\Delta t=1$, matches with the needed charge $s_n^k$.


The total power consumed by the PEVs getting charged at time interval $t$ appears in the power flow equations of the distribution network as $p_{i,t}^{EV}$ where $i$ is the power system node hosting the aggregator EV charging facility. The aforementioned coupling is captured by
\begin{equation}
p_{i,t}^{EV}=\sum_{n=1}^{N_k} P_{n,t}^{(k)},~\forall t.
\end{equation}
Additionally, the maximum number of PEVs getting charged at a time interval is limited by the maximum number of available charging spots $\overline{\mathsf{N}}$
\begin{equation}
    \mathbf{1}^\top\mathbf{D}^{(k)}\leq\overline{\mathsf{N}}\mathbf{1}^\top
\end{equation}


\subsection{Objective Function}
At the beginning of $k$-th interval, the aggregator would try to prepare a schedule over the next $T_k$ intervals that maximizes the profit. Let $\mathcal{N}_1^k$ and $\mathcal{N}_2^k$ be the set of PEV owners willing to pay the charging prices $C_1$ (\$/kWh) and $C_2$ (\$/kWh) respectively. Thus, the total number of new PEVs is $n_k=|\mathcal{N}_1^k\cup\mathcal{N}_2^k|$. Based on the prepared schedule, the aggregator would have to pay the electricity prices to the utility. With $\mathbf{c}_{e}\in\mathbb{R}^{T_k\times1}$ representing the electricity prices for the next $T_k$ instances, the anticipated electricity cost is given by $\mathbf{1}^\top\mathbf{P}^{(k)}\mathbf{c}_{e}$. Therefore, the scheduling problem solved by the aggregator at the start of $k$-th time interval can be formulated as an MILP

\begin{align}\label{eq:P1}
\min_{\mathbf{D}^{(k)},\mathbf{p}^{(k)},\mathbf{u}}&~\mathbf{1}^\top\mathbf{P}^{(k)}\mathbf{c}_{e}-\sum_{n\in\mathcal{N}_1^k} C_1u_ns_n^k-\sum_{n\in\mathcal{N}_2^k} C_2u_ns_n^k\tag{P1}\\
\mathrm{s.to}~&~ (1)-(9)\notag
\end{align}

Problem~\eqref{eq:P1} if feasible yields a possible schedule such that the PEVs selected for establishing a charging contract ${(u_n=1)}$ receive their requested charging $s_n^k$ within the agreed time $a_n^k$. We next delineate the algorithm for periodically solving \eqref{eq:P1} in a moving horizon basis while ensuring feasibility and profitability.

\subsection{Moving time horizon updates}
In a realistic setup for local aggregators, the number of PEVs $n_k$, arriving at time $k$ is not known a priori. Therefore, an optimal schedule can not be generated at the start of the day. Rather, based on the initial $n_1$, an initial candidate schedule may be prepared, which is subsequently updated. In detail, at the start of any interval $k$, the aggregator shall solve \eqref{eq:P1} and obtain a schedule for the next $T_k$ intervals; implement the schedule for $k$-th interval; discard the remaining schedule and resolve \eqref{eq:P1} at the begining of $k+1$-th interval.

The proposed framework gurantees that the chrging commitments $(s_n,a_n)$ are fulfilled for PEVs with contract ($u_n=1$). To see this, note that for the first instance $t=1$, contracts are established for PEVs for which there exists a feasible schedule $\mathbf{P}^{(1)}$ over the next $T$ intervals. Thus, after implementing the first interval, the remaining $T-1$ columns of $\mathbf{P}^{(1)}$ are still a candidate solution to \eqref{eq:P1} for $t=2$ with no new contracts established; hence guaranteeing feasibility of \eqref{eq:P1}. Similarly, since the truncated schedule from $\mathbf{P}^{(1)}$ is still feasible, any new schedule generated as $\mathbf{P}^{(2)}$ must yield a higher profit by optimality. Continuing the argument for all intervals, it is evident that the novel approach of enforcing \eqref{eq:un} and \eqref{eq:chargetime} in a moving horizon basis guarantees fulfillment of commitment towards PEV owners while maximizing profit.

The initialization and updation of parameters $s_n^k$'s and $a_n^k$'s are explained next. When a new PEV arrives, its charging energy requirement is computed as
$$s_n=\frac{(\mathsf{{SOC}_{plugout}}-\mathsf{SOC_{plugin}}) \times \mathsf{Bcap_n}}{\eta \times \Delta t}$$
where $\mathsf{SOC_{plugout}}$ is the final SOC of an PEV to be reached and $\mathsf{SOC_{plugin}}$ is the SOC at the PEV's time of arrival. $\mathsf{Bcap_n}$ is the battery capacity of $n^{th}$ PEV. Once the total energy requirement $s_n$ is computed, the guranteed \emph{time of return} $a_n$ may be computed based on the price option $C_1$ or $C_2$ opted by the PEV owner. In detail, let $P_{C_1}$ and $P_{C_2}$ be the average charging power for the two cost options with $P_{C_1}>P_{C_2}$ for $C_1>C_2$. Then the time of return is given by
$$a_n^k=\ceil[Bigg]{s_n/P_{C_j}},~\forall~n\in\mathcal{N}_j^k.$$

Once the optimal schedule is obtained for $k^{th}$ time, the charging power requirement and time availability of the selected $M_k$ PEVs are updated for the $(k+1)^{th}$ time interval.
\begin{equation}
\begin{aligned}
    &s_n=s_n^{(k)}-P^{(k)}(n,1),\quad &\forall n\textrm{ with }u_n^{(k)}=1\\
    &a_n=a_n^{(k)}-1,\quad &\forall n\textrm{ with }u_n^{(k)}=1\\
    &c_n=c_n^{(k)},\quad &\forall n\textrm{ with }u_n^{(k)}=1
\end{aligned}
\label{eq:update}
\end{equation}


\section{Numerical Tests}
The proposed approach is tested on a single phase IEEE 13-node feeder with nominal voltage of 4.16 kV~[]. To obtain a load curve for a 24 hr period, real-world residential demands from Pecan street Data was used [16]. In detail, 15-min based data for 300 houses was taken from Pecan Street; summed and normalized to obtain an hourly demand profile. Since the Pecan Street data-set does not provide reactive power demands, a power factor of $0.9$ lagging was considered to generate normalized reactive power demands. Next, the normalized profile was scaled by the spot-load data form the 13-node feeder. The PEV charging station was assumed to be located at node~$5$ of the feeder. Problem~\eqref{eq:P1} was solved using YALMIP and Gurobi, on a 2.7 GHz Intel Core i5 computer with 8GB RAM~[17],~[18].

\begin{figure}[t]
\centering
\includegraphics[width=1\linewidth]{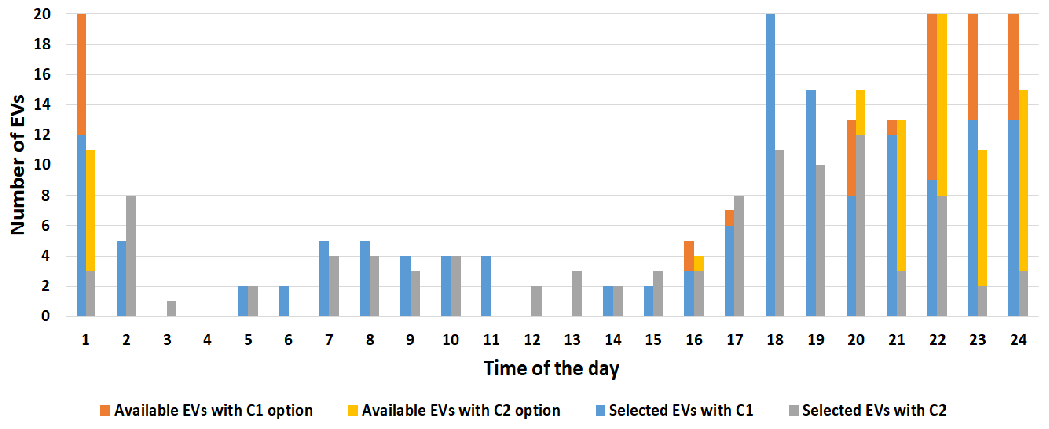}
\caption{Number of available and selected PEVs over 24 hours}\label{Fig:}
\end{figure}

For a random arrival of PEVs opting the two price options $C_1$ and $C_2$, Fig. 2 shows the PEV selection process over 24 hours (12 AM- 11:59 PM). It can be seen that, at in the initial off peak hours, almost all the PEVs are being selected to be charged within the committed time. However, that is not the case during evening peak hours as the base load increases and possible time for charging decreases.

\begin{figure}[t]
\centering
\includegraphics[width=1\linewidth]{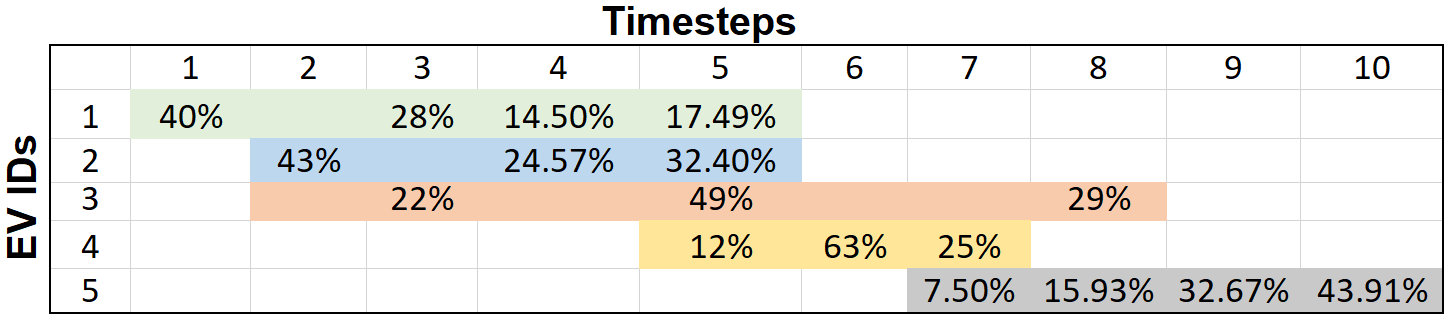}
\caption{PEV charging power allocation}\label{Fig:}
\end{figure}

The novelty of the algorithm is the guarantee the aggregator provides in terms of charging time. The charging power distribution for a subset of PEVs over 10 time steps is shown in the Fig. 3. The shaded area for each PEV depicts its availability for charging and the percentage of charging completed in every time step. All PEVs are shown to receive 100$\%$ of their charging requirement. Further, it can be observed that the charging rate is non-uniform and charging is discontinuous, as anticipated.

Next the variation of active power load of the EV charging station in response to price and power system load variations is shown in Fig.~4. All quantities are normalized for clarity, wherein the EV charging load is normalized with the transformer rating at node $5$. It may be observed that, the charging station demand diminishes at high price period to minimize cost, while it complements the network-load to alleviate over/under voltages. Finally, to analyze scalability, $100$ instances of \eqref{eq:P1} were solved for random arrival of PEVs over a 24 hr period. The total time for solving \eqref{eq:P1} was found to be in the range~$[12.7,~21.4]$~sec with median at $16.8$~sec. 

\begin{figure}[t]
\centering
\includegraphics[scale=0.145]{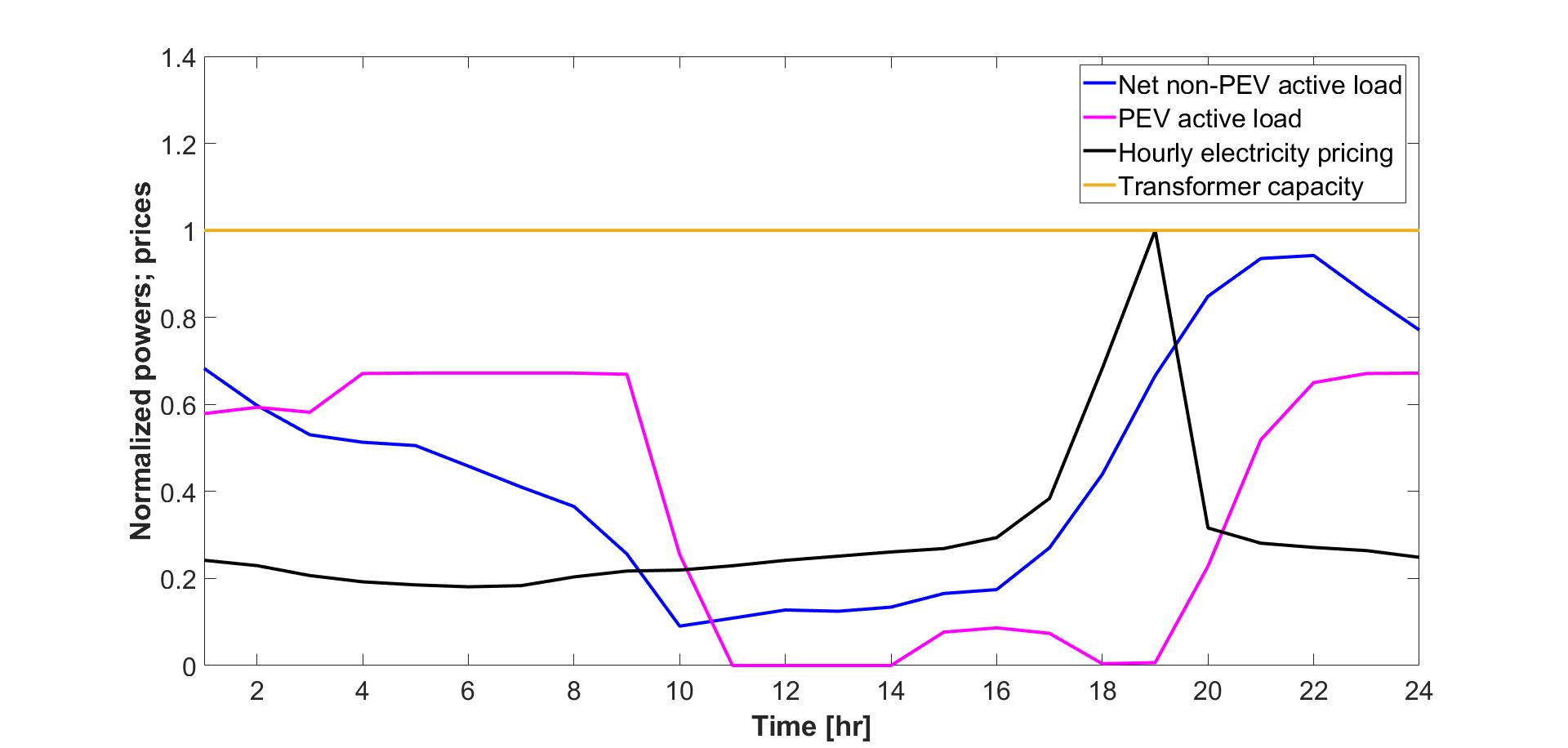}
\caption{Intraday profile of power system load in response to EV station load and electricity prices}\label{Fig:}
\end{figure}

\section{CONCLUSION}
The developed formulation gives optimal charging scheduling of incoming PEVs, considers system constraints and also maximizes the number of PEVs charged at a real-time scenario. The simulation results validate the proposed MILP formulation for a moving time horizon. It can be scaled down to smaller time intervals for detailed system framework. Also, it can be extended for multiple aggregator charging decisions in a medium-sized power system network with higher PEV penetration. Although simulation did not involve distributed energy resources, it was seen in the problem formulation that it can be well accommodated in the design. As part of future work, this model is supposed to be improvised to allow vehicle to grid power flow at peak hours along with capacity market constraints to delve deeper into the economy of the PEV charging market.

\end{document}